\documentclass{PoS}
\usepackage{wrapfig}
\usepackage{url}
\newcommand\Rad{\Phi_{\rm rad}}
\newcommand\PSb{\Phi_B}
\def\lq{\left[}
\def\rq{\right]}

\title{Latest theory developments for top pair production, generators and showering}

\ShortTitle{Latest theory developments for top pair production, generators and showering}

\author{\speaker{Emanuele Re}\\~\\
  Rudolf Peierls Centre for Theoretical Physics,\\
  University of Oxford, 1 Keble Road, Oxford, UK\\~\vspace{-0.2cm}\\
  Laboratoire d'Annecy-le-Vieux de Physique Theorique (LAPTh),\\
  9 chemin de Bellevue, Annecy-le-Vieux, France\thanks{current address}\\
  E-mail: \email{emanuele.re@lapth.cnrs.fr}}

\abstract{I summarize the state of the art of cross-section
  computations and of available simulation tools for top-quark
  pair production in hadron collisions. Particular emphasis is put on
  recent theory developments relevant for LHC phenomenology.}

\FullConference{8th International Workshop on Top Quark
  Physics\\ 14-18 September, 2015\\ Ischia, Italy}

\begin{document}

\section{Introduction}
\vspace{-0.05cm}
Top Physics is a central part of the LHC Physics programme, as shown
for instance by the variety of topics and results presented at this
conference. The large value of the cross sections to produce top
quarks at the LHC allows an experimentally accurate extraction of its
properties. On the other hand, the final state arising from its decay
products makes top-quark production processes a background for several
BSM searches. For these reasons it is extremely important to have
predictions for total and differential cross sections that are precise
enough to match the experimental accuracy. It is also crucial to model
as accurately as possible subleading effects by including higher-order
corrections into event generator programs.

In this paper I review the state of the art of computations and
simulation tools for top-quark pair production, that is the dominant
mechanism for producing top quarks at the LHC. In
section~\ref{sec:partonic} I focus on results for total and differential
cross sections at parton level, whereas section~\ref{sec:MC} reviews the
current research activity devoted to the improvement of simulation
tools.

\section{Recent improvements in the computation of total and differential cross sections}
\label{sec:partonic}
\vspace{-0.05cm}
In this section I will first summarize recent highlights in the
computation of the $t\bar{t}$ total and differential cross sections in
QCD (sec.~\ref{sec:fixed}-\ref{sec:resum}). In the last part
(sec.~\ref{sec:differential}) I give the state of the art for fully
differential predictions where top quark decay products are included.

\subsection{Fixed order results}
\label{sec:fixed}
\vspace{-0.05cm}
\begin{wrapfigure}{r}{0.35\columnwidth}
\vspace{-0.3cm}
\centerline{\includegraphics[width=0.4\textwidth]{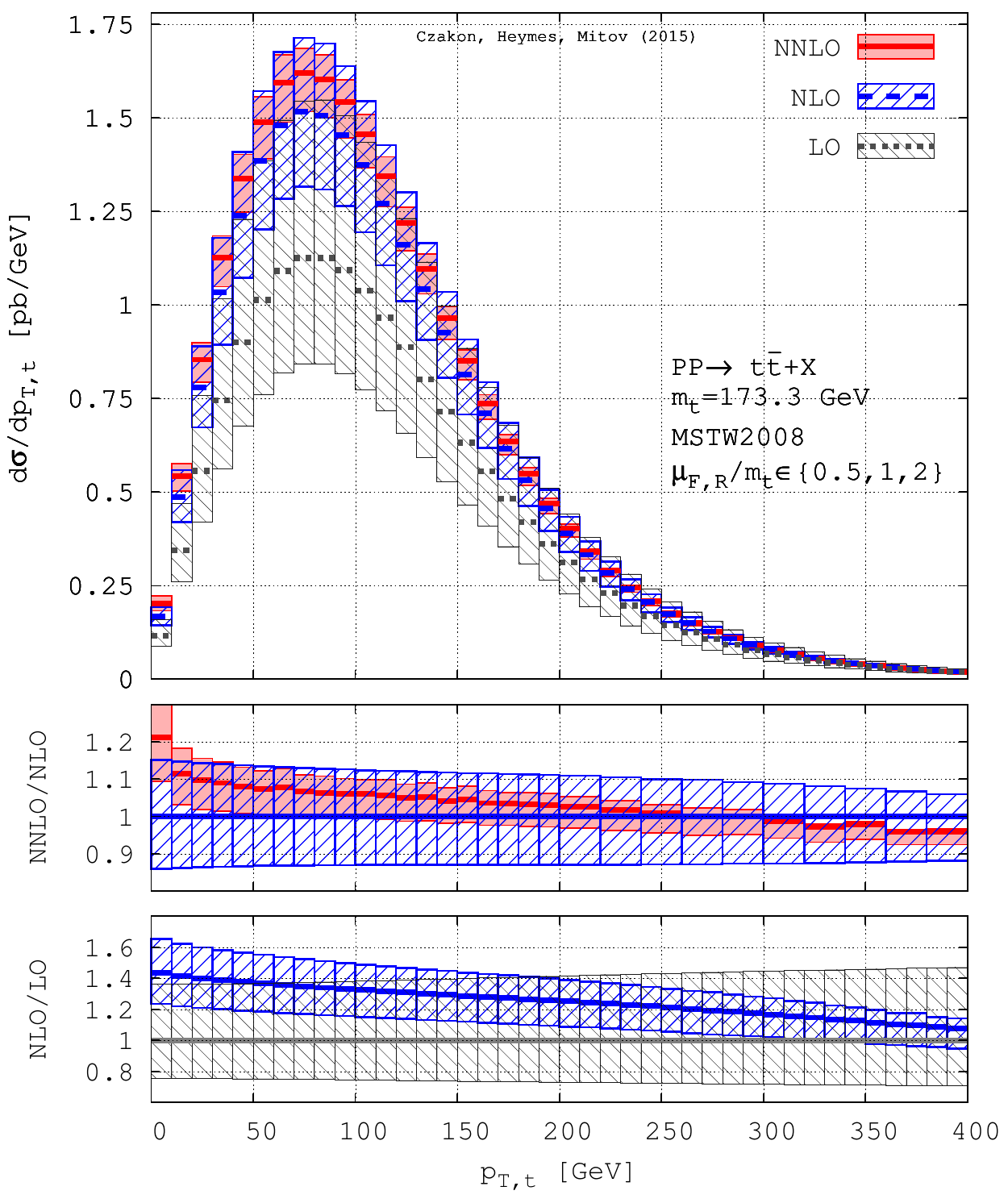}}
\vspace{-0.1cm}
\caption{The top quark transverse momentum at LO, NLO and NNLO, and
  the associated K-factors (LHC, $\sqrt{S}=8$ TeV). Figure from
  ref.~\cite{Czakon:2015owf}.}
\label{fig:Czakon15}
\vspace{-0.0cm}
\end{wrapfigure}

A landmark result for top quark Physics at the LHC was obtained by the
authors of ref.~\cite{Czakon:2013goa}, that have computed for the first
time the exact next-to-next-to-leading order (NNLO) QCD corrections to
the total inclusive cross section $\sigma_{tot}$, using the
subtraction scheme presented in ref.~\cite{Czakon:2010td}. These
results show a fairly good
convergence of the perturbative series, with good overlap among
uncertainty bands when going from one order to the next one. The
agreement between theory and data is also remarkably good, as shown in
several talks at this conference.

Differential distributions for $t\bar{t}$ production are now also
known at NNLO in QCD. They were first published for the forward
backward top asymmetry in ref.~\cite{Czakon:2014xsa}, and, very
recently, for other (more exclusive)
observables~\cite{Czakon:2015owf}. The importance of the latter
results can not be underestimated. NNLO/NLO differential K factors are
generally not flat, as illustrated for instance the plot in
fig.~\ref{fig:Czakon15}, where it is shown that NNLO corrections
change sizeably the shape of the top quark $p_T$ spectrum.\footnote{
  It has to be noticed though, that a fixed renormalization and
  factorization scale is used in these results.} In the tail region
several studies by ATLAS and CMS have shown an unsatisfactory
agreement between data and NLO-matched event generators, with the
latter overshooting the measured cross section.  By including NNLO
corrections, the agreement is significantly improved, as also shown in
ref.~\cite{Czakon:2015owf}.  An important refinement of these
differential results will be its combination with the NNLO
differential computation of the top decay, that was performed in
refs.~\cite{Gao:2012ja,Brucherseifer:2013iv}.

For completeness I also recall that significant progress towards the
computation of NNLO QCD corrections to total and differential
$t\bar{t}$ cross sections was also made by other
groups~\cite{Abelof:2015lna,Bonciani:2015sha}. For instance the cross
section $d\sigma/dp_{T,t\bar{t}}$ for the transverse momentum of the
top pair is now known in the small $p_T$ limit at the
next-to-next-to-leading logarithmic
order~\cite{Li:2013mia,Catani:2014qha}: steps towards a generalization
of the ``$q_T$-subtraction'' method to perform a NNLO computation with
colored final states have started to be taken~\cite{Bonciani:2015sha},
thanks to the availability of these results.

\subsection{All-order results}
\label{sec:resum}
\vspace{-0.05cm} When the $t\bar{t}$ system is produced just above
threshold, effects due to multiple soft-gluon radiation become
relevant. In this phase-space region large logarithmically-enhanced
corrections of the form $\alpha_s^n \log^m X_s$ appear, where $X_s$ is
a kinematic function that vanishes when the $t\bar{t}$ system is
produced exactly at threshold. Thanks to factorization properties of
phase space and production matrix elements in the soft region (the
latter being not trivial for $t\bar{t}$ due to color flow), the
perturbative expansion can be reorganized suitably, by computing
systematically the log-enhanced terms, and resumming them to all
orders.

Several groups performed soft-gluon threshold resummation at
next-to-next-to-leading logarithmic (NNLL)
accuracy~\cite{Czakon:2009zw,Kidonakis:2010dk,Ahrens:2010zv,Beneke:2011mq,Cacciari:2011hy},
using different formalisms and different functional forms for $X_s$,
and in one case also resumming Coulomb
effects~\cite{Beneke:2011mq}. NNLL resummation has also been matched
to the exact NNLO computation mentioned in sec.~\ref{sec:fixed}: this
accuracy represent the state of the art for $\sigma_{tot}$, and is
available in public codes as {\ttfamily TOP++}~\cite{Czakon:2013goa}
and {\ttfamily TOPiXS}~\cite{Beneke:2012wb}.


Resummed results can also be used to guess the size of the first
unknown term at fixed order: until the exact NNLO result was
available, approximate NNLO results have been computed by several
groups.  Nowadays the natural development is to try and compute
$\sigma_{tot}$ at approximate N$^3$LO. This has been achieved, using
different approaches, in ref.~\cite{Kidonakis:2014isa}, as well as in
ref.~\cite{Muselli:2015kba} (in the latter paper, high-energy
resummation was used together with threshold resummation).

\subsection{Fully differential computations, including top-quark decays}
\label{sec:differential}
\vspace{-0.05cm}
Since top quarks can not be detected, $\sigma_{tot}$ and differential
distributions for stable top quarks can not be directly compared with
data without some extrapolation. In this respect, differential results
that include also the final state decay products are of fundamental
importance, and are also a central ingredient for the latest
developments of modern event generators, to be discussed in
sec.~\ref{sec:MC}. The more accurate fixed-order results for
describing the fully exclusive final state arising from ``top pair''
production and decay are those presented, a while ago, in
refs.~\cite{Denner:2012yc,Bevilacqua:2010qb,Heinrich:2013qaa,Frederix:2013gra,Cascioli:2013wga}. Despite
there being differences among the computations, all these papers
essentially contain fully differential NLO results for the
process\footnote{Throughout the document with this notation it is
  implied that leptonic $W$ decays are included.}  $pp\to W^+
W^- b\bar{b}$, where all offshellness effects as well as interference
among double-, singly- and non-resonant diagrams are taken into
account exactly up to NLO in QCD. A remarkable result was recently
obtained in ref.~\cite{Bevilacqua:2015qha}, where the same effects
were also computed at NLO for the process $pp\to W^+ W^- b\bar{b} +1$
jet.

It is worth stressing that two of the aforementioned computations
(namely~\cite{Frederix:2013gra} and~\cite{Cascioli:2013wga}) were
performed in the ``4-flavor'' scheme, \emph{i.e.} the $b$ quarks are
considered massive, hence results are finite also in the limit of
vanishing $p_T$ for one (or both) $b$ quarks (as well as for
unresolved $b$-tagged jets). Therefore these computations allow a
clean comparison among theory predictions (at NLO) and data for the
experimental signature traditionally called ``single-top $Wt$''
production: since arbitrary cuts can be safely placed on $b$ jets,
consistent results can be obtained for the same final state detected
experimentally (2 $b$-tagged jets vs.~\emph{e.g.} 1 $b$-tag and 1
$b$-veto), thereby avoiding the ambiguities intrinsically present when
a separation between top-pair and $Wt$ production is attempted
theoretically. An experimental analysis dedicated to a comparison
among these computations and data would certainly be very valuable.

\section{Recent improvements in event generators}
\label{sec:MC}
\vspace{-0.05cm} Fully exclusive Monte Carlo event generators based on
parton-shower algorithms are used ubiquitously in experimental
searches, hence their importance for LHC phenomenology doesn't need to
be stressed in this short review. A substantial step to improve the
accuracy of these simulation tools has been made more than a decade
ago, when methods to consistently match NLO QCD computations with
parton showers algorithms were devised (NLO+PS). Since then, an
enormous progress took place in this field: all important processes of
the type $pp\to t\bar{t} +X$ can be simulated at NLO+PS, thanks to the
developments of fully- or partially-automated
frameworks~\cite{Alioli:2010xd,Alwall:2014hca,Hoeche:2011fd,Platzer:2011bc,Kardos:2011qa}. Among
them, {\ttfamily MadGraph5\_aMC@NLO}~\cite{Alwall:2014hca} deserves to
be specially mentioned, being it the only one which is currently fully
automated in the strictest sense.
For other frameworks, if a specific process of interest is not
publicly available, its NLO+PS simulation can be obtained with minor
efforts, by linking against external codes (typically, to obtain
1-loop amplitudes with large multiplicity). Nowadays this can be done
straightforwardly, using standard interfaces developed specifically
for this purpose~\cite{Binoth:2010xt,Alioli:2013nda}.

There are currently two very active research topics in the community
of Monte Carlo developers that are relevant for top pair production at
hadron colliders: the consistent inclusion of offshellness effects in
presence of intermediate resonances decaying into colored particles,
and the merging of NLO+PS simulations for different jet
multiplicities. I will review them in turn, with particular emphasis
on the former.

\subsection{Simulation of $pp\to W^+W^-b\bar{b}$ at NLO+PS accuracy}
\vspace{-0.05cm}
Before turning to the explanation of the theoretical issues presently
addressed by the community, I want to recall that a major application
where a NLO+PS simulation of of $pp\to W^+W^-b\bar{b}$ can have an
impact is in the determination of the top mass, at least for the
techniques where the kinematics of visible particles from top-decay is
used for this purpose (see~\cite{Corcella:2015kth} for a recent
review).

The problem with the simulation of $W^+W^-b\bar{b}$ production can be
stated as follows: unless special care is taken, when NLO correction
to the decay are included in NLO+PS tools, the intermediate top-quark
virtuality is not preserved. If this happens, non-physical distortion
can potentially show up in kinematic distributions.
Although there are issues also with the {\ttfamily MC@NLO} method (see
for instance~\cite{streview}), from this point onward I'll focus on
the {\ttfamily POWHEG} approach.

Before entering into details, a remark is due: it is
legitimate to ask whether the
issues discussed below are really relevant for practical purposes,
especially because NLO+PS results for the $W^+W^-b\bar{b}$ final state
(with offshellness and interference effects) were obtained with
{\ttfamily PowHel} in ref.~\cite{Garzelli:2014dka}, and no particular
problems were noticed by the authors. The definitive answer can only
be given by developing more refined tools and performing careful
comparisons among them and against older approaches.

The {\ttfamily POWHEG} ``master formula'' to generate a resolved
emission reads
\begin{equation}\nonumber
  d\sigma = d\PSb d\Rad \bar{B}(\PSb) \frac{R(\PSb,\Rad)}{B(\PSb)}\exp{\lq-\int \frac{R(\PSb,\Rad)}{B(\PSb)}d\Rad\rq}\,.
\end{equation}
The concept of ``underlying Born'' phase space ($\PSb$) is central in
{\ttfamily POWHEG}, and we assume the reader to be familiar with it:
once a point in $\PSb$ is picked, according to the weight $\bar{B}$,
the hardest emission (a point in $(\PSb,\Rad)$) is generated according
to the {\ttfamily POWHEG} Sudakov. The mapping $\PSb\to(\PSb,\Rad)$ is
the same as the one used to perform the subtraction of singularities
present in the $R$ term contained within the $\bar{B}$ function, and
it depends on the singular region at hand~\cite{Frixione:2007vw}. Two
problems are present:\footnote{Extended explanations can be found in
  refs.~\cite{Campbell:2014kua,Jezo:2015aia}.}
\begin{enumerate}
\item In the standard {\ttfamily POWHEG BOX} algorithm, the phase
  space region associated to final-state gluon emission off the
  $b$-quark would be handled by a mapping that, in general, does not
  preserve the virtuality of the intermediate resonance, \emph{i.e.}
  $m^2_{Wb}(\PSb)\ne m^2_{Wbg}(\PSb,\Rad)$. The problem is manifest:
  unless $m^2_{bg} \ll \Gamma_t E_{bg}$, $R$ and $B$ will not be on
  the resonance peak at the same time, hence the ratio $R/B$ can
  become large when $R$ is on peak and $B$ is not, yielding a
  ``Sudakov suppression'' that is spurious, since the $(\PSb,\Rad)$
  kinematics would be far from the true QCD
  singularity. Quantitatively, one expects the mass profile of the
  $b$-jet to be distorted when $m^2_{jet}\sim E_b \Gamma_t$.
\item A further problem can arise during the parton-showering stage:
  from the second emission onward, the shower should be instructed to
  preserve the mass of the resonances. This could be done easily if
  there was an unique mechanism to ``assign'' the radiation to a given
  resonance. For processes where interference is present, no obvious
  choice is possible.
\end{enumerate}
  
An intermediate solution to the previous issues was presented in
ref.~\cite{Campbell:2014kua}, where a fully consistent NLO+PS
simulation for $W^+W^-b\bar{b}$ production in the narrow-width limit
was obtained. Offshellenss and interference effects were implemented
in an approximate way, as follows:
\begin{itemize}
\item[a.] By using matrix elements in the narrow-width limit, real and
  virtual corrections for production and decay can be clearly
  separated,~\emph{i.e.} no interference arises. This also allows for
  a non-ambiguous ``resonance assignment'' for final-state
  radiation.
\item[b.] For radiation in the decay, $\Rad$ is generated by first
  boosting momenta in the resonance rest-frame. In this way, the
  intermediate virtuality is the same for $\Rad$ and $(\PSb,\Rad)$.
\item[c.] The phase space integration, and the event generation, spans
  also over the off-shell regions. A projection onto an on-shell
  kinematics, tested extensively, allows to use the NLO on-shell
  amplitudes (computed in refs.~\cite{Campbell:2012uf,Badger:2011yu}).
\item[d.] From the off-shell phase space, a reweighting of the
  $\bar{B}$ function is performed using the LO exact results (where
  finite width and non-double-resonant diagrams are fully included).
\end{itemize}

\begin{wrapfigure}{r}{0.4\columnwidth}
\vspace{-0.3cm}
\centerline{\includegraphics[width=0.4\textwidth]{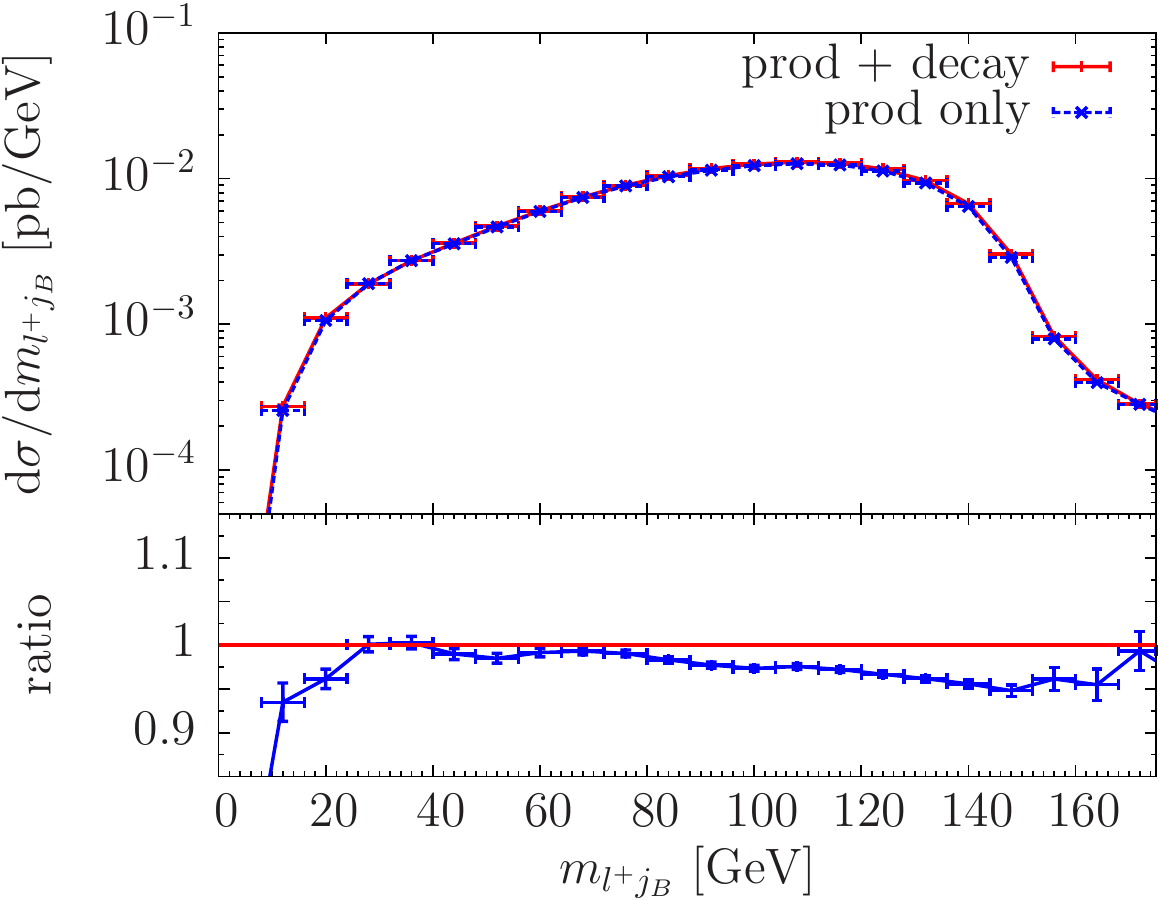}}
\vspace{-0.1cm}
\caption{Invariant mass distribution of the charged lepton and
  $b$-flavored jet at NLO+PS including correction for production only
  or for production and decay (LHC, $\sqrt{S}=8$ TeV), with the
  method described in ref.~\cite{Campbell:2014kua}. Figure adapted
  from ref.~\cite{Campbell:2014kua}.}
\label{fig:CENR14}
\vspace{0.0cm}
\end{wrapfigure}
The above expedients allow the construction of a NLO+PS generator
where the theoretical problems mentioned above are solved, and
offshellness effects are included approximately. A further issue was
addressed: by default in {\ttfamily POWHEG} only the hardest emission
is generated. However, for the $t\bar{t}$ process, emissions from
decay are rarely the hardest, hence they would be dealt with by the
shower most of the time, despite the previous improvements. In
ref.~\cite{Campbell:2014kua} a procedure to keep, at the same time,
the initial state radiation as well as those from decaying resonances
was implemented, to alleviate the aforementioned issue. An example of
the results
is illustrated in fig.~\ref{fig:CENR14}, where the impact of the new
{\ttfamily POWHEG BOX} generator is shown on an ``endpoint''
observable typically used to extract $m_t$.

Finally, I want to mention that, shortly after the conference, further
substantial progress was presented in ref.~\cite{Jezo:2015aia}. A
method to handle exactly the complete matrix elements also at NLO was
developed, by partially using some of the improvements in
ref.~\cite{Campbell:2014kua} but also generalizing substantially the
partition of phase space into singular regions and the associated
subtraction scheme. Although results were published only for
single-top, the method is fully general, and its application to
$W^+W^-b\bar{b}$ is in progress~\cite{bb4lpaper}.

\subsection{Multijet merging at NLO}
\vspace{-0.05cm}
At large collision energies, a significant fraction of $t\bar{t}$
events is produced in association with one or more jets. At times a
tool describing several jet multiplicities in a single event sample is
needed. A typical example is when ``$H_T$'' variables are used, as is
often the case in BSM searches.

The CKKW-L and MLM-merging methods succesfully address this issue at
LO. Since this accuracy will become a limiting factor for precision
studies, it is desirable to extend these methods to NLO (``NLOPS
multijet merging''). Reaching such accuracy is a non-trivial
theoretical challenge, since it requires a detailed understanding of
the interplay among resummation and fixed order effects in NLO+PS
simulations. Several approaches were proposed in the literature over
the last three years. In the context of top-pair production, so far
results were published only using the {\ttfamily
  MEPS@NLO}~\cite{Hoeche:2012yf} and {\ttfamily
  FxFx}~\cite{Frederix:2012ps} merging
methods.\footnote{Phenomenological studies performed by the original
  authors can also be found in
  refs.~\cite{Hoeche:2013mua,Hoeche:2014qda} and
  ref.~\cite{Alwall:2014hca}. Moreover a thorough comparison where
  other approaches are also included will be presented in the
  proceedings of the 2015 ``Physics at TeV Colliders''
  workshop~\cite{LHproceedings}.}  The measurement of QCD activity in
$t\bar{t}$ events will allow to test these tools against data in ``SM
dominated'' regions, thereby providing a robust assessment of the
accuracy that can be assumed when they are used for BSM searches.


\end{document}